# Improving MMA judging with consensus scoring: A Statistical analysis of MMA bouts from 2003 to 2023

Vincent Berthet

University of Lorraine, 2LPN, F-54000 Nancy, France
Sorbonne Economics Centre, CNRS UMR 8174, Paris, France

vincent.berthet@univ-lorraine.fr

January 3, 2024

## Abstract

Boxing and MMA have a longstanding issue with judging, as evidenced by frequent controversial decisions. Like boxing, MMA bouts are scored following the 10-Point Must System, by which judges score each round individually. In the present study, we compare the performance of two methods of round scores aggregation: the standard method (aggregation over rounds and then over judges) and an alternative method known as consensus scoring (aggregation over judges and then over rounds). For that purpose, we conducted a statistical analysis of 4,129 MMA bouts that went to decision between 2003 and 2023. Our findings indicate that standard and consensus scoring yield the same result in 97.53% of the bouts. Noteworthy, this percentage may be underestimated, as judges may not always score each round individually; instead, they may balance their scores across rounds. Regarding the bouts in which the two methods disagree, the outcome of consensus scoring aligns more with the opinion of the fans (48.96%) than that of standard scoring (43.75%). This results from the fact that consensus scoring can counteract the impact of an incorrect score (especially a 10-8 score) given by a judge in a specific round. From a cost-benefit perspective, we argue that state commissions should consider consensus scoring in MMA as an alternative to standard scoring as it can avoid controversial decisions while it does not imply a major change in the judging system.

# 1. Introduction

Combat sports such as boxing and MMA belong to the category of sports that are scored subjectively by judges. Like boxing, MMA bouts are scored by three judges ringside following the 10-Point Must Scoring System according to which 10 points must be awarded to the winner of the round and nine points or less must be awarded to the loser, except for an even round, which is scored 10-10. According to the Unified Rules of MMA, "A 10-9 Round in MMA is where one combatant wins the round by a close margin" (most common case), "A 10-8 Round in MMA is where one fighter wins the round by a large margin" (rare case), while a "A 10-7 Round in MMA is when a fighter completely overwhelms their opponent" (extremely rare case). Most of MMA bouts are scheduled in three rounds of five minutes while certain bouts (those with a title on the line) are scheduled in five rounds. The total scores of the judges are used to decide the winner if the fight goes the distance.

While there is evidence that MMA judges mostly follow the evaluation criteria provided to them (Feldman, 2020), MMA judges' scores can be unreliable and inconsistent. The UFC Fight Night 227 (also known as Noche UFC) took place on September 16, 2023 at T-Mobile Arena in Las Vegas. The main event was the UFC Women's Flyweight Championship rematch between current champion Alexa Grasso and former champion Valentina Shevchenko. Grasso retained her title in a controversial split draw. Judge Sal D'Amato gave the fight 48-47 to Shevchenko, while judge Junichiro Kamijo had it 48-47 for Grasso. The third judge was Mike Bell, who scored it 47-47, which resulted in the split draw. Afterward, it was revealed that Bell gave the fifth round to Grasso with a 10-8, which led to the 47-47 scorecard (see Table 1). This single 10-8 score cost the fight to Shevchenko. If Mike Bell had scored the fifth round 10-9 for Grasso, like the two other judges, Shevchenko would have won the fight by split decision.[1]

**Table 1.** Official judges' scorecards from the bout between Alexa Grasso and Valentina Shevchenko 2 at Noche UFC on September 16, 2023. In each round, the consensus score is the score given by at least two out of the three judges.

|    | Mike Bell | | Sal D'Amato | | Junichiro Kamijo | | Consensus score | |
|----|-----------|-----------|-------------|-----------|------------------|-----------|-----------------|-----------|
|    | Grasso | Shevchenko | Grasso | Shevchenko | Grasso | Shevchenko | Grasso | Shevchenko |
| R1 | 9  | 10 | 9  | 10 | 9  | 10 | 9  | 10 |
| R2 | 10 | 9  | 10 | 9  | 10 | 9  | 10 | 9  |
| R3 | 9  | 10 | 9  | 10 | 9  | 10 | 9  | 10 |
| R4 | 9  | 10 | 9  | 10 | 10 | 9  | 9  | 10 |
| R5 | 10 | 8  | 10 | 9  | 10 | 9  | 10 | 9  |
|    | 47 | 47 | 47 | 48 | 48 | 47 | 47 | 48 |

The rematch between Grasso and Shevchenko highlights the issue of *noise* in MMA and boxing judges – specifically, the occasional low inter-judge agreement or inter-rater reliability (e.g., Stallings & Gillmore, 1972). It is known that noise can impact judges in various contexts (Kahneman et al., 2021). For instance, Clancy et al. (1981) asked 208 federal judges to determine the appropriate sentences for the same 16 cases. The cases were described by the characteristics of the offense (e.g., robbery or fraud, violent or not) and of the defendant (e.g., young vs. old, repeat vs. first-time offender). While the mean sentence was 7 years, the average difference between the sentences that two randomly chosen judges gave for the same crime was more than 3.5 years. Thus, the level of inter-agreement between these federal judges was quite low.

---

[1] The week following this controversial decision, the Nevada State Athletic Commission (NSAC) hold a training session for all licensed judges that specifically focused on the criteria for a 10-8 round.

Various factors may lower the inter-judge agreement in boxing and MMA. Judges are located in different locations ringside and thus they do not see the action from the same view.[2] Moreover, a recent study has shown that MMA judges have different individual preferences regarding fighting actions. For instance, some judges deem missed significant head strikes as positive actions, while others believed they are negative (Holmes et al., 2024). It should be noted that judges may exhibit various biases such as home advantage (e.g., see Balmer et al., 2005, for evidence in boxing), bias toward larger betting favorites, those with insurmountable leads, and the fighter who won the previous round (Gift, 2017). However, if all judges have the same biases, the inter-judge agreement would still be high.

The 10-Point Must System serves as a scoring mechanism in boxing and MMA, by which judges score each round individually. Subsequently, round scores can be aggregated through various methods employing a majority rule to determine the overall winner of the fight. The standard procedure involves adding up the round scores assigned by each judge and applying the majority rule to these cumulative totals (aggregation over rounds and then over judges). An alternative method is to consider the majority (or consensus) score for each individual round (i.e., the score given by at least two out of the three judges) and sum these round scores (aggregation over judges and then over rounds) (see Table 1 for an example). In boxing, such a method is known as *consensus scoring*, which was developed and submitted to the National Association of Attorneys General Boxing Task Force by Dr. Ralph S. Levine of Stanford University.

Under consensus scoring, the consensus score for a given round is calculated as follows. In cases where at least two judges have the same score, that score is the consensus score (e.g., if the three judges' round scores were 10-9, 10-9, and 10-8, the consensus score would be 10-9). Otherwise, the judge's score which most favors Fighter A, as well as the score which most favors Fighter B are dropped and the remaining score is the consensus score. For example, if the judges' scores were 10-9, 10-10, 9-10, the consensus score (median score) would be 10-10 as the 10-9 and 9-10 scores would be dropped.

A method closely resembling consensus scoring has been proposed in boxing by the former New Jersey State Athletic Control Board (NJACB) Commissioner, Larry Hazzard, Sr., under the name of the "10-Point Majority System". Figure 1 shows how the majority score for a given round is calculated.

---

[2] An often-cited example illustrating the impact of judges' positioning is the UFC Welterweight Championship bout between Georges St-Pierre and Johny Hendricks at UFC 167 that took place on November 16, 2013. The three judges scored the five rounds identically, except for round 1. In this round, judge Glenn Trowbridge scored it 10-9 for Hendricks, while the other two judges scored it 10-9 for St-Pierre. It is believed that Trowbridge awarded the round to Hendricks based on a specific action he observed from his position.

**Figure 1.** The 10-Point Majority System (adapted from Algranati & Cork, 2000).

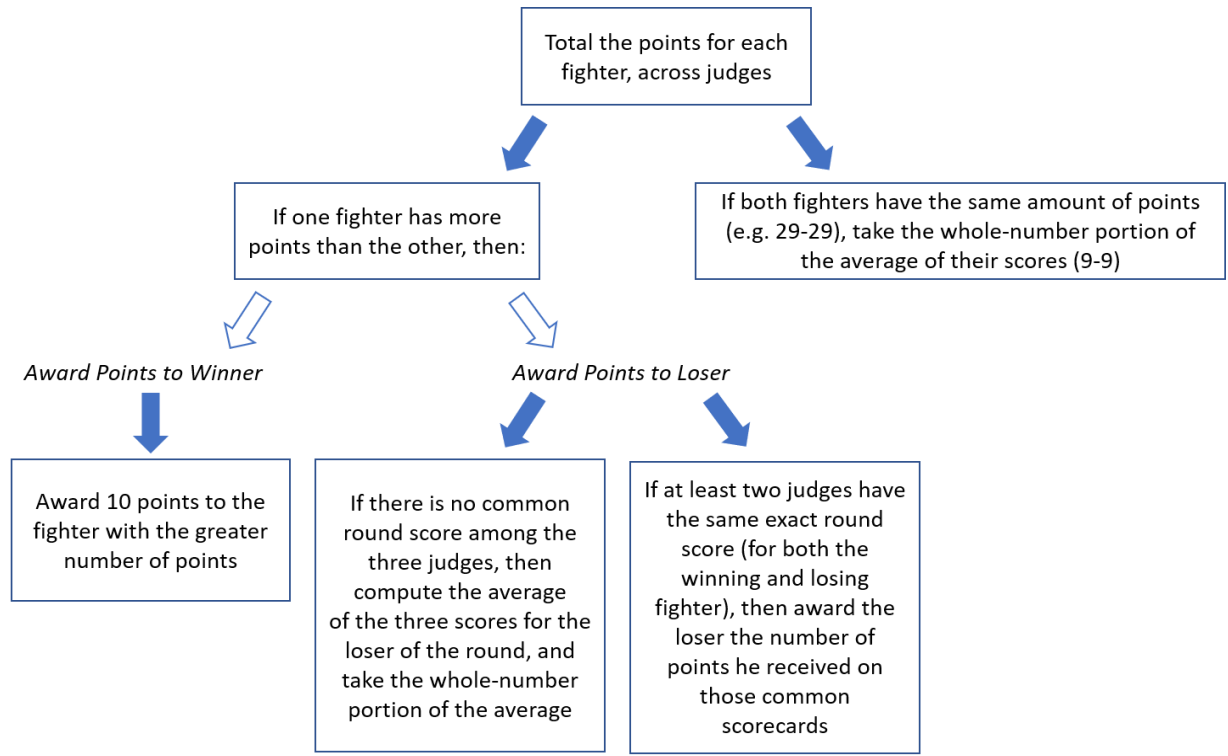

The consensus scoring system and the 10-Point Majority System have in common that a score aggregated over judges is formed on a round-by-round basis. When two judges agree on how a round is to be scored, the consensus score and the majority score are identical. These two methods may differ when it comes to rounds in which there is no agreement among the judges (i.e., the three judges have three different scores). For instance, if the three judges' scores were 10-8, 10-10, and 9-10, the consensus score would be 10-10 while the majority score under the 10-Point Majority System would be 10-9 (Fighter A has a total of 29, Fighter B has a total of 28; Fighter A is awarded 10 points, Fighter B is awarded the integer part of (8+10+10)/3 which is 9). However, our statistical analysis of over 4,000 MMA bouts shows that at least two judges have the same score in virtually every round (see Results section). Thus, we will consider consensus scoring in the present study.

A major problem with standard scoring is that it is sensitive to rounds miscalled by a single judge. By contrast, consensus scoring can counteract the impact of an incorrect score given by a judge in a specific round. Indeed, if the miscall of a round is caused by chance, it is unlikely that another judge will make the same mistake in the same round. Figure 2 shows an example of such a phenomenon in the bout between Claudio Silva and Leon Edwards at UFC Fight Night 56 on November 8, 2014. Silva won the fight in a controversial split decision. Roy Silbert scored the second round 10-9 for Silva, while the two other judges and 95% of the fans scored it 10-9 for Edwards. Thus, Roy Silbert arguably miscalled the second round. Under consensus scoring, the consensus score in round 2 would have been 10-9 for Edwards, and Edwards would have won the fight 29-28. Likewise, Mike Bell arguably miscalled the fifth round 10-8 for Grasso in her rematch against Shevchenko, which led to a controversial split draw. Under consensus scoring, the consensus score in round 5 would have been 10-9 for Grasso, and Shevchenko would have won the fight 48-47 (see Table 1).

**Figure 2.** Example of a bout in which standard scoring resulted in a controversial split decision due to a miscalled round, while consensus scoring would have identified the "correct" winner (screenshot made from mmadecisions).

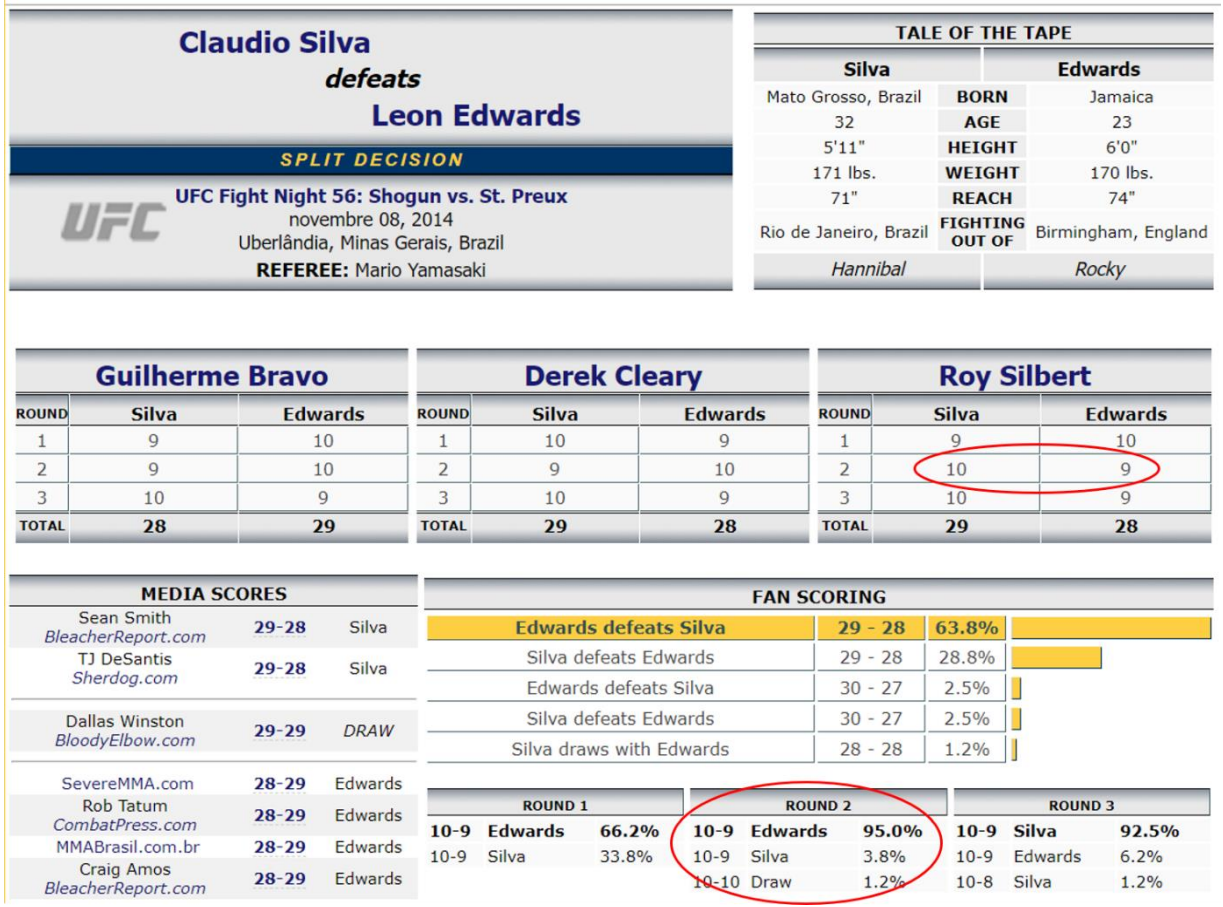

It follows that consensus scoring should also lower the impact of a single biased judge on the final outcome, provided that the other judges do not exhibit the same bias.[3] Baumann and Singleton (2023) have recently conducted simulations showing that deciding boxing bouts by the majority score rule (consensus scoring), compared with deciding by the majority judges rule (standard scoring), makes it less likely that a biased judge sways the outcome of a bout. The aim of the present study is to compare the performance of standard and consensus scoring on actual MMA bouts that went to decision.

## 2. Methods

### 2.1 Database

The data were collected from mmadecisions.com. The dataset includes all MMA bouts that went to decision from September 26, 2003 (UFC 44: Undisputed) to October 7, 2023 (UFC on ESPN+ 87 and Bellator 300), which represents a total of 19,883 rounds. As can be seen in Table 2, most of rounds are from UFC (52.05%) and Bellator (21.50%) bouts. For each round, we collected the judges' and the fans' scores. Indeed, fans can submit their scorecards on mmadecisions.com.

---

[3] A frequently cited example of such biased judging in professional boxing is the world championship bout that took place on March 13, 1999, between Lennox Lewis and Evander Holyfield in New York City (Lee et al., 2002).

**Table 2.** Number of rounds as a function of organization.

| Organization | N rounds |
|---|---|
| Bellator | 4274 |
| Birmingham | 21 |
| Boxing | 382 |
| CW | 1175 |
| EFN | 3 |
| Invicta | 731 |
| KSW | 680 |
| PFL | 752 |
| ShoMMA | 198 |
| Strikeforce | 464 |
| UFC | 10350 |
| WEC | 367 |
| WSOF | 486 |

### 2.2 Data cleaning

We removed the rounds from boxing bouts (382 rounds), the rounds in which the name of the judges was missing (1,193 rounds), the rounds in which the scores were missing (5,269 rounds), the bouts in which the round scores were not available for all rounds (5 bouts), and the rounds in which there was no majority score (10 rounds). The final sample included 13,024 rounds, corresponding to a total of 4,129 bouts.

## 3. Results

### 3.1 Inter-judge agreement

The percentage of rounds with a unanimous score is 71.68%, which means that inter-judge agreement is quite high. Among the rounds without a unanimous score (1,169 rounds), two judges have the same score (majority score) in virtually every round, except for ten rounds (0.86%) (see Appendix).

### 3.2 Comparison of standard and consensus scoring

Table 3 shows that standard and consensus scoring yield the same result in 97.53% (4,027 out of 4,129) of the bouts in the dataset.

**Table 3.** Comparison of the outcome of 4,129 MMA bouts under standard and consensus scoring. Note that in the mmadecisions database, fighter A is always the winner of the fight.

|  |  | **Consensus scoring decision** | | | |
|---|---|---|---|---|---|
|  |  | Fighter A | Fighter B | Draw | Total |
| **Standard scoring decision** | Fighter A | 3985 | 70 | 11 | 4066 |
|  | Draw | 12 | 9 | 42 | 63 |
|  | Total | 3997 | 79 | 53 | 4129 |

We then analyzed the performance of the two methods on the remaining 2.47% of fights (N=102) with regard to the fans' opinion. Indeed, we reasoned that when they disagree, the method that is the most consistent with the majority of public opinion should be viewed as better. One could rightly point out that the fans' opinion cannot be viewed as the gold standard for several reasons. First and foremost, fans are not professional judges and it is likely that some of them are not aware of the official judging criteria in MMA. Second, the scores submitted by fans on mmadecisions can be affected by various biases such as loser bias (fans who disagree with the outcome are more likely to submit a scorecard than those who are content with the decision) and geographical bias (sometimes certain countries generate more traffic than others, which could lead to a bias in fan voting). However, we note that fans and standard scoring agree in 81.27% of the bouts, which means that the fans' scores on mmadecisions are largely plausible. In the 102 bouts where standard and consensus scoring disagree, we found that the outcome under consensus scoring aligns more closely with the fans' opinion (48.96%) than does that under standard scoring (43.75%).

Noteworthy, our analysis relies on the assumption that the judges would have scored the rounds the same way, regardless of the scoring system (standard or consensus) in use. Otherwise, the level of agreement between the two methods would be underestimated. In fact, we hypothesize that under standard scoring, judges may not always score each round individually as they may balance their scores across rounds. Figure 3 shows an example of such a phenomenon in the bout between Tae Hyun Bang and Leo Kuntz at UFC Fight Night 79 on November 28, 2015. Charlie Keech scored the second round 10-9 for Bang while the two other judges and 85% of the fans scored it 10-9 for Kuntz. Keech might have corrected this mistake by scoring the third round 10-9 for Kuntz, while another judge and 80% of the fans scored it 10-9 for Bang.

**Figure 3.** Example of a bout in which a judge (Charlie Keech) may have balanced his scores across rounds 2 and 3. (screenshot made from mmadecisions).

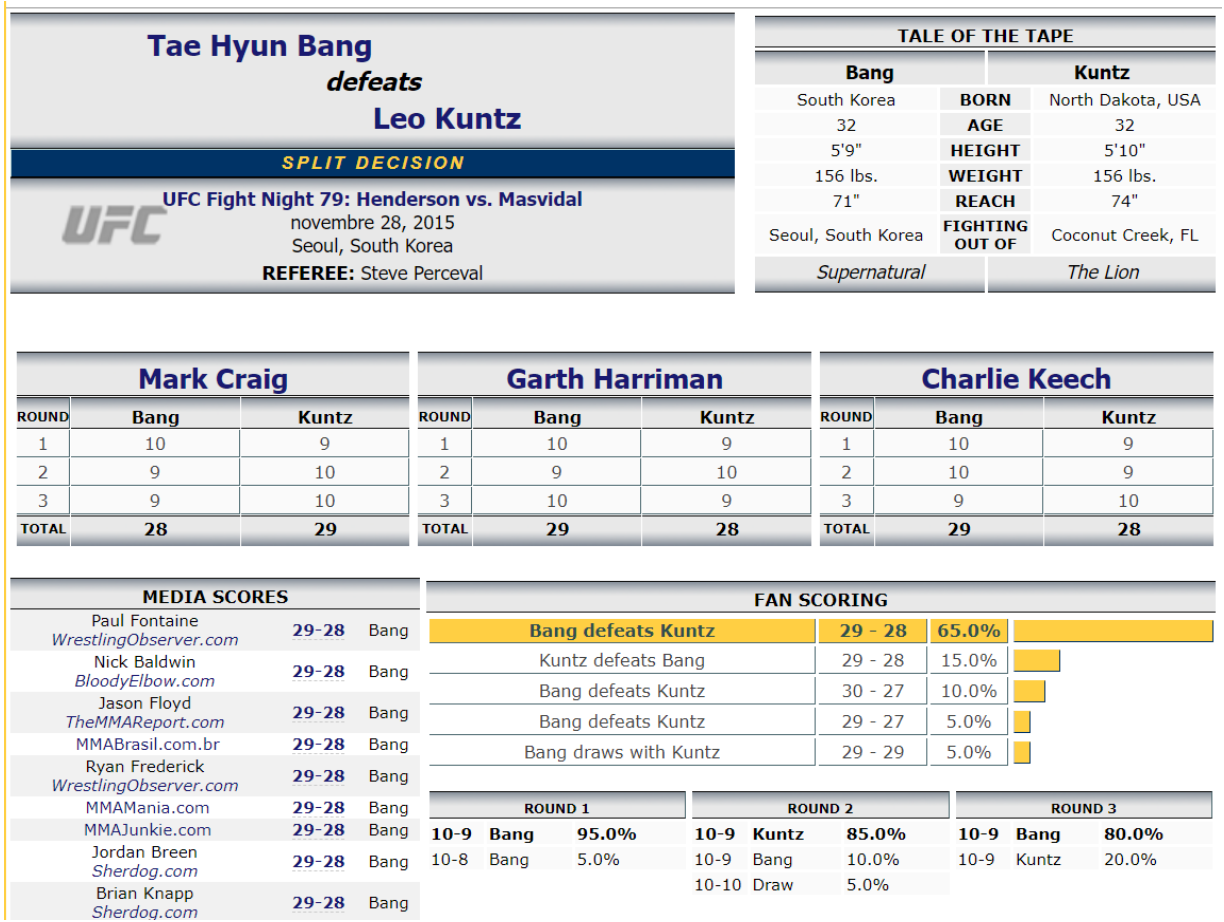

Estimating the number of bouts in which at least one judge balanced his/her scores across rounds among the 102 bouts where standard and consensus scoring disagree would be a difficult task. However, such a phenomenon may occur, and we note that the true level of agreement between the two methods is likely to be higher than 97.53% if judges actually score each round individually.

The bout detailed in Figure 3 highlights the fact that under consensus scoring, judges should be instructed and trained to not balance their scores across rounds. In this fight, consensus scoring would wrongly identify Kuntz as the winner (standard scoring and the public identify Bang as the winner) because one judge (Charlie Keech) may have balanced his scores across rounds 2 and 3. Had that judge scored round 3 individually (despite having miscalled round 2), consensus scoring would correctly identify Bang as the winner.

## 4. Discussion

The main result of our study is that standard and consensus scoring yield the same result in 4,027 out of 4,129 MMA bouts that went to decision between 2003 and 2023 (97.53%) (see Algranati & Cork, 2000, for similar findings in boxing). One could argue that such a high percentage of agreement between the methods favors the status quo. However, the 2.47% of cases where the two methods disagree include high-stakes fights (e.g., the UFC Championship rematch between Alexa Grasso and Valentina Shevchenko). Therefore, determining the fairest method of scoring bouts remains an important issue.

Regarding the bouts in which standard and consensus scoring disagree, the outcome under consensus scoring aligns more with the opinion of the fans (48.96%) than does that under standard scoring (43.75%). This results from the fact that consensus scoring can counteract the impact of an incorrect score (especially a 10-8 score) given by a judge in a specific round.

Following the controversial decision in the March 13, 1999 championship fight between Lennox Lewis and Evander Holyfield, the National Association of Attorneys General Boxing Task Force highlighted consensus scoring as a "new system of scoring to maximize fair and accurate results" and emphasized the need for state commissions to "evaluate and consider such alternative systems". History repeats itself: following the controversial decision in the September 16, 2023 UFC championship fight between Alexa Grasso and Valentina Shevchenko, we once again emphasize the need for state commissions to consider consensus scoring in MMA.

It is important to note that consensus scoring is no miracle solution to the issues surrounding MMA judging. Within the 10-Point Must System, this method can counteract the impact of incorrect scores from a single judge, but not incorrect scores from two judges. For instance, among the 10 most controversial decisions of 2023 listed by mmadecisions[4], consensus scoring would have identified a different winner in only one of them. The reason for this is that such controversial decisions involve questionable scoring by two judges.

Finally, the analysis conducted in the present study relies on the assumption that the judges would have scored the rounds the same way, regardless of the scoring system in use. Such an assumption is questionable. Another way to compare different scoring systems is to implement them on the same sample of future MMA bouts over a significant period of time (several years) while ensuring that judges genuinely score each round individually. Such a study could first be conducted in amateur MMA leagues.

---

[4] http://mmadecisions.com/blog/130/The-10-Most-Disputed-Decisions-of-2023

# Appendix

**Table.** The ten rounds in the mmadecisions database (13,024 rounds between 2003 and 2023) in which the three judges had three different scores.

| Event | Round | Judge 1 | | Judge 2 | | Judge 3 | |
|---|---|---|---|---|---|---|---|
| UFC Fight Night 32 | 1 | Douglas Crosby | | Marcos Rosales | | Otto Torriero | |
| | | Ferreira | Sarafian | Ferreira | Sarafian | Ferreira | Sarafian |
| | | 10 | 10 | 9 | 10 | 10 | 9 |
| CWFC 59 | 1 | Ben Cartlidge | | David Lethaby | | Andy Sledge | |
| | | Hill | Moore | Hill | Moore | Hill | Moore |
| | | 9 | 10 | 10 | 9 | 10 | 10 |
| UFC 159 | 2 | Eric Colón | | Michael Depasquale Jr. | | Jose Tabora | |
| | | Saint Preux | Villante | Saint Preux | Villante | Saint Preux | Villante |
| | | 10 | 9 | 9 | 10 | 10 | 10 |
| KSW 22 | 3 | Leszek Pawlega | | Wojslaw Rysiewski | | Cezary Wojciechowski | |
| | | Bedorf | Thompson | Bedorf | Thompson | Bedorf | Thompson |
| | | 10 | 9 | 10 | 10 | 9 | 10 |
| KSW 29 | 2 | Ben Cartlidge | | Radoslaw Grela | | Cezary Wojciechowski | |
| | | Khalidov | Cooper | Khalidov | Cooper | Khalidov | Cooper |
| | | 9 | 10 | 10 | 9 | 10 | 10 |
| KSW 29 | 3 | Szymon Bonkowski | | Ben Cartlidge | | Radoslaw Grela | |
| | | Silva | Strus | Silva | Strus | Silva | Strus |
| | | 10 | 10 | 10 | 9 | 9 | 10 |
| KSW 29 | 1 | Szymon Bonkowski | | Irena Preiss | | Cezary Wojciechowski | |
| | | Reljic | Narkun | Reljic | Narkun | Reljic | Narkun |
| | | 10 | 10 | 9 | 10 | 10 | 9 |
| Invicta FC 36 | 3 | Kevin Champion | | Greg DiVilbiss | | Steven Graham | |
| | | Harpe | Norton | Harpe | Norton | Harpe | Norton |
| | | 10 | 9 | 9 | 10 | 10 | 8 |
| Invicta | 3 | Kevin Champion | | Greg DiVilbiss | | Steven Graham | |
| | | Stoliarenko | Verzosa | Stoliarenko | Verzosa | Stoliarenko | Verzosa |
| | | 9 | 10 | 10 | 8 | 10 | 9 |
| UFC on ESPN+ 25 | 1 | Derek Cleary | | Sal D'Amato | | Ester Lopez | |
| | | Dvalishvili | Kenney | Dvalishvili | Kenney | Dvalishvili | Kenney |
| | | 9 | 10 | 10 | 9 | 10 | 8 |